\documentclass[]{paper_template}

\usepackage{amsmath}
\usepackage{booktabs}
\usepackage{listings}

\newcommand{\methodname}{TTSonic}

\title{TTSonic: A Unified Framework for Composable and Controllable Speech Generation}

\author[1,*,\dagger]{Zhedong Zhang}
\author[1,*,\dagger]{Kaiwen Zhou}
\author[1]{Junyan Wu}
\author[1]{Shengqiang Li}
\author[1]{Xuran Chi}
\author[1]{Xingui~Zeng}
\author[1]{Rui Yan}
\author[1,\ddagger]{Xu Chen}
\affiliation[1]{ByteDance Douyin Content Group}
\emails{\texttt{\{zhangzhedong.zzd,\ zhoukaiwen.0124,\ junyanwu,\ 
  shengqiangli.01,\ chixuran,\ zengxingui,\ yanrui.001,\ 
  chenxu.722\}@bytedance.com}}
\contribution[*]{Equal contribution.}
\contribution[\dagger]{Work was done during internship at ByteDance Douyin Content Group.}
\contribution[\ddagger]{Corresponding author.}

\renewcommand{\paperlogo}{\includegraphics[height=1cm]{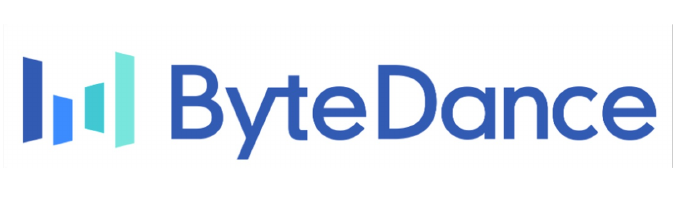}}

\abstract{
Modern text-to-speech (TTS) systems can synthesize natural and expressive
speech, yet a practical speech foundation model must go beyond naturalness to
offer stable quality, precise controllability, and a broad set of composable
capabilities within a unified framework.
We present \methodname{}, a unified speech generation model that supports
zero-shot TTS, voice cloning, prosody cloning, instruction following, voice conversion, and cloning from noisy reference
speech. Rather than relying on separate task-specific models, \methodname{}
accommodates diverse conditioning signals, including text, natural-language
instructions, reference speech, and text-described timbre, within the same
model framework.
Specifically, we develop a semantic--acoustic disentangled speech tokenizer
through ASR-supervised semantic learning and reconstruction-guided generative
refinement. Serving as the representational foundation of our unified
framework, the tokenizer enables the Text-to-Semantic autoregressive model to
focus on semantic and prosodic planning, while the acoustic flow-matching model
specializes in high-quality speech rendering under diverse acoustic and timbre
conditions.
Building on this framework, we incorporate CoT-guided instruction-following
planning and Text2Timbre voice design to support controllable speech
generation.
Experiments across zero-shot synthesis, multilingual and cross-lingual TTS,
instruction following, voice design, voice conversion, and noisy-reference
cloning demonstrate the competitive performance and broad applicability of
\methodname{}.
}

\demopage{\url{https://bytedance.github.io/TTSonic/}}

\begin{document}
\maketitle

\section{Introduction}
\label{sec:introduction}

Text-to-speech (TTS) synthesis has progressed from producing intelligible
speech to generating natural, expressive, and speaker-faithful audio
\citep{shen2018natural,renfastspeech}. Modern systems trained on large-scale
multi-speaker corpora can synthesize high-quality speech from only a few
seconds of reference audio, while recent advances further enable multilingual
generation, long-form synthesis, and natural-language control
\citep{du2024cosyvoice2,du2025cosyvoice3,hu2026qwen3,zhou2026voxcpm2}. As TTS becomes
an enabling technology for conversational agents\citep{roy2026personaplex, xie2025fireredtts}, content creation\citep{zhang2026instructdubber,cong2026cosyncdit}, live
streaming, and interactive services\citep{wang2025streammel}, the objective is no longer merely to
produce speech that is natural and intelligible. Instead, a practical TTS system
must offer stable synthesis quality, precise controllability, robust voice
cloning from imperfect references, and a unified interface for deciding
\emph{what is said}, \emph{how it is said}, and \emph{who says it}.

Despite these advances, integrating diverse speech generation capabilities
into a single model remains an open challenge. Many existing speech generation
models are optimized for a narrow operating mode,
such as zero-shot TTS, prompt-based voice cloning, cloning from noisy prompt, instruction-following
synthesis, text-guided voice design, or voice conversion. Extending a system
from one capability to another often requires separately trained models,
making it difficult to maintain consistent quality and predictable control
across applications. 
A useful speech foundation model should therefore accept diverse
conditioning signals and compose multiple generation capabilities instead of
treating each task as an isolated problem.

To this end, we introduce \methodname{}, a unified and composable speech
generation framework that adapts to diverse tasks and available conditions by
recomposing shared modules: the speech tokenizer, Text-to-Semantic Speech LM,
acoustic flow-matching decoder.
Specifically, we train the \methodname{} Speech Tokenizer through a two-stage
learning process. In the first stage, ASR-supervised semantic pre-training
encourages the encoder to extract semantic and prosodic representations that
are disentangled from speaker-dependent acoustic characteristics. In the
second stage, we freeze the semantic encoder and refine the FSQ mapping with
reconstruction feedback from a pre-warmed acoustic branch, together with an
auxiliary speech-understanding objective. This generative refinement preserves
the semantic structure learned in the first stage while making the resulting
discrete tokens better suited for downstream speech generation.
Built on this semantic representation, the text-to-semantic autoregressive
language model performs long-context semantic and prosodic planning from text
and optional natural-language instructions. The acoustic flow-matching model
then renders accurate and high-quality speech conditioned on user-provided
reference audio or voice design instruction. This factorized design makes
\methodname{} composable within a unified framework: task-dependent generation
paths are formed by selecting, adding, or reusing semantic, prosodic,
reference-speech, and timbre conditions, rather than training separate
end-to-end models for individual capabilities. The shared representation and
acoustic generation framework assign complementary roles to these conditions,
allowing semantic content, speaking style, and speaker identity to be
controlled and recombined across diverse speech generation tasks.

Under this unified generation paradigm, \methodname{} supports zero-shot TTS,
voice cloning, prosody cloning, controllable cloning, voice conversion, and
cloning from noisy reference speech. A separate Text2Timbre model generates
speaker embeddings from textual descriptions and supplies them to the acoustic
model, enabling flexible voice design and timbre switching without modifying
the unified generation model.
We further employ CoT-guided style planning as an explicit intermediate
representation for instruction-following control.

\begin{table}[h!]
    \centering
    \scriptsize
    \setlength{\tabcolsep}{4pt}
    \renewcommand{\arraystretch}{1.15}
    \caption{\textbf{Generation paradigms supported by \methodname{}.}}
    \label{tab:generation-mode-inputs}
    \resizebox{0.9\linewidth}{!}{
        \begin{tabular}{@{}lll@{}}
            \toprule
            \textbf{Generation Mode}
            & \textbf{Model}
            & \textbf{Generation Paradigm} \\
            \midrule
            Speech Enhancement & codec,fm & Noise speech $\rightarrow$ Clean speech\\
            Voice Conversion & codec,fm & Content Speech, Timbre Speech $\rightarrow$ Conversion Speech\\
            Voice Cloning & llm,fm & Text, Prompt Speech $\rightarrow$ Target Speech\\
            Prosody Cloning & codec,llm,fm & Text, Prompt Speech, Prosody Speech $\rightarrow$ Target Speech\\
            Voice Design (TTS) & llm,fm,vd & Text, Voice Description $\rightarrow$ Target Speech\\
            Voice Design (Voice Conversion) & codec,fm,vd & Content Speech, Voice Description $\rightarrow$ Target Speech\\
            Controllable Cloning & llm,fm & Text, Style Description, Prompt Speech $\rightarrow$ Target Speech\\
            Instruct Following & llm,fm,vd & Text, Instruction $\rightarrow$ Target Speech\\
            \bottomrule
        \end{tabular}
    }
\end{table}

As summarized in Table~\ref{tab:generation-mode-inputs}, \methodname{} supports
diverse speech generation and transformation tasks through flexible
compositions of the speech tokenizer, Text-to-Semantic Speech LM, acoustic
flow-matching model, and the separate Text2Timbre module.

Our contributions can be summarized as follows:
\begin{itemize}
    \item We propose \methodname{}, a unified, composable, and controllable
    speech generation framework that supports a broad set of capabilities,
    including zero-shot TTS, voice cloning, prosody cloning, controllable
    cloning, voice conversion, and robust cloning from noisy reference speech.
    These capabilities arise from task-dependent compositions of shared modules
    rather than task-specific end-to-end models. A separate Text2Timbre module
    further enables flexible voice design and timbre switching.

    \item We propose a semantic--acoustic disentangled speech tokenizer through
    ASR-supervised semantic learning and reconstruction-guided generative
    refinement, serving as the representational foundation of our unified model.
    The resulting discrete representations preserve semantic and prosodic
    information while disentangling speaker-dependent acoustic characteristics.

    \item Extensive experiments across zero-shot synthesis, multilingual and
    cross-lingual TTS, instruction following, voice design, voice conversion,
    and noisy-reference cloning demonstrate the competitive performance and
    broad applicability of \methodname{}.
\end{itemize}

\section{Method}
\label{sec:method}

\begin{figure}[!t]
    \centering
    \includegraphics[width=0.8\linewidth]{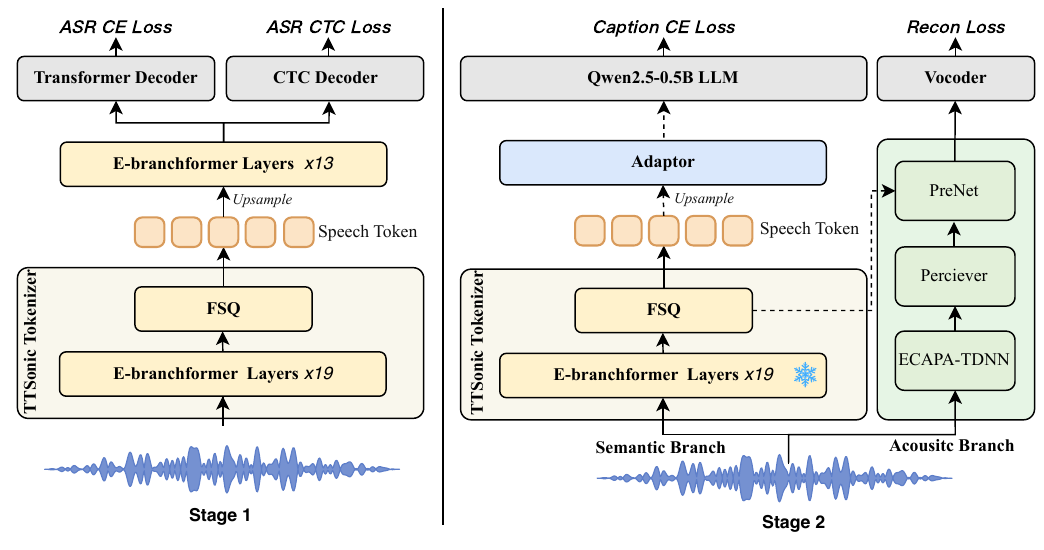}
    \caption{\textbf{Training pipeline of the TTSonic Speech Tokenizer.}
    Stage I inserts the FSQ bottleneck after the first 19 E-Branchformer blocks,
    then upsamples the quantized features and processes them with 13 additional
    blocks under encoder--decoder ASR, CTC, and consistency supervision. Stage II
    freezes the first 19 blocks and updates only the FSQ mapping, using speech
    reconstruction and speech understanding for multi-task supervised learning.}
    \label{fig:codec-training}
\end{figure}


\subsection{TTSonic Speech Tokenizer}
\label{sec:method-codec}

\paragraph{Architecture.}
The TTSonic Speech Tokenizer converts speech into a compact sequence of
semantic speech tokens for subsequent language-model training. Given a 16-kHz
waveform $x$, an 80-dimensional filter-bank frontend followed by the first 19
blocks of a 512-dimensional E-Branchformer encoder~\citep{kim2022ebranchformer}
produces continuous semantic representations
\begin{equation}
    H = E_{\mathrm{sem}}^{1:19}(x).
    \label{eq:codec-semantic-encoder}
\end{equation}
The encoder reduces the temporal resolution to 25 Hz. We quantize $H$ with
finite scalar quantization (FSQ)~\citep{mentzer2024fsq}. The quantizer first
projects every 512-dimensional frame into a five-dimensional latent, whose
scalar levels are $[8,8,8,8,4]$, and maps the quantized latent back to 512
dimensions:
\begin{equation}
    s_t = Q_{\mathrm{sem}}(h_t),
    \qquad
    z_t = \mathrm{Dec}_{Q}(s_t).
    \label{eq:codec-semantic-token}
\end{equation}
This construction defines a single implicit codebook of $16{,}384$ entries and
produces one discrete speech token every 40 ms. The deployed tokenizer consists
of the 19 E-Branchformer blocks and the FSQ mapping. The downstream
text-to-speech system consumes this frame-level token sequence as the semantic
representation predicted by the language model.

\paragraph{Two-stage training.}
Figure~\ref{fig:codec-training} summarizes the two-stage tokenizer training
pipeline. The training strategy first builds a stable semantic encoder and then
uses multi-task supervised learning to optimize the discrete token mapping used
by the generative model.

\textbf{Stage I: ASR semantic pre-training with CTC regularization.}
We first train the tokenizer in an encoder--decoder automatic speech recognition
framework, augmented with connectionist temporal classification (CTC)
loss~\citep{graves2006ctc} and consistency
regularization~\citep{yao2025crctc}. As shown in
Figure~\ref{fig:codec-training}, the FSQ bottleneck is inserted after the first
19 E-Branchformer blocks (layer index 18). The quantized features are then
upsampled and processed by 13 additional E-Branchformer blocks before being fed
to the Transformer ASR decoder and CTC decoder. These 13 blocks form the Stage-I
supervision stack and are not part of the deployed tokenizer. The ASR decoder
objective provides strong recognition supervision, while the CTC loss directly
constrains the encoder-side frame-level representations. Placing the FSQ
bottleneck at this intermediate-late position encourages the discrete tokens to
retain linguistic and prosodic information. In ablations over the FSQ insertion
point, moving
the bottleneck to an intermediate-late encoder layer improves semantic feature
extraction, while excessively late insertion becomes suboptimal. This suggests
that the upper E-Branchformer blocks provide cleaner abstractions for
tokenization up to a suitable depth, after which further delaying
discretization may reduce the usefulness of the discrete representation. In
addition, paired stochastic
views of an utterance are regularized by the frame-level divergence between
their posterior distributions. The objective is
\begin{equation}
    \mathcal{L}_{\mathrm{ASR}}
    =
    \mathcal{L}_{\mathrm{dec}}
    +\lambda_{\mathrm{ctc}}\mathcal{L}_{\mathrm{CTC}}
    +
    \lambda_{\mathrm{cr}}
    \mathcal{L}_{\mathrm{cons}}.
    \label{eq:codec-asr}
\end{equation}
This combined supervision prevents the autoregressive decoder from absorbing
all linguistic modeling and encourages the E-Branchformer semantic encoder to
learn a relatively clean representation of phonetic, lexical, and prosodic
structure.

\begin{figure}[!t]
    \centering
    \includegraphics[width=\linewidth]{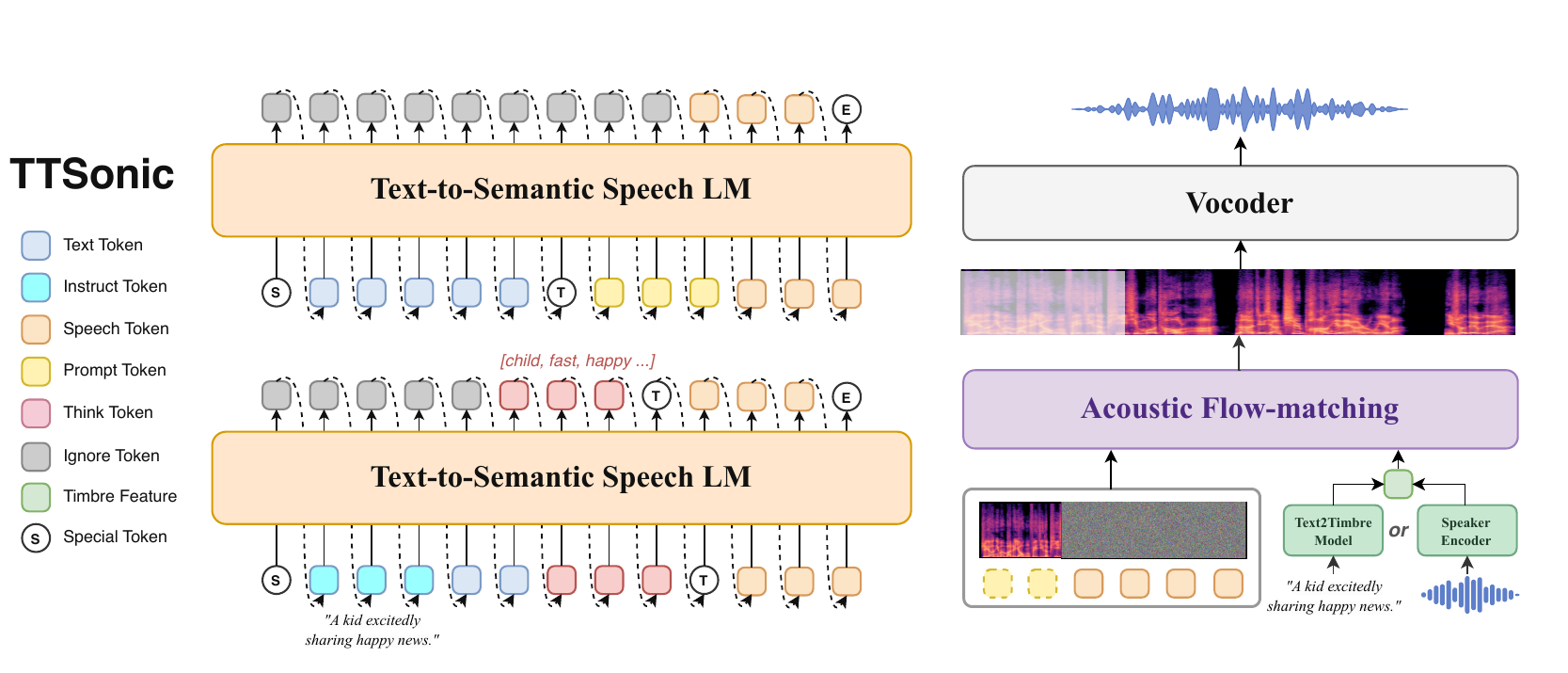}
    \caption{\textbf{Overview of \methodname{}.}
    \methodname{} consists of three main components: the TTSonic Speech
    Tokenizer, a Text-to-Semantic Speech LM, and an acoustic flow-matching
    model. The Text-to-Semantic Speech LM supports both standard voice cloning
    and CoT-guided instruction following, while the flow-matching model can be
    conditioned on speaker features extracted from reference audio or on
    timbre embeddings designed by the Text2Timbre model.}
    \label{fig:ttsonic-overview}
\end{figure}

\textbf{Stage II: multi-task supervised learning.}
We initialize the 19-block semantic encoder from Stage I and keep it frozen
throughout the second stage. Only the FSQ module on top of this frozen encoder
is updated; the additional 13-block ASR supervision stack used in Stage I is no
longer part of the tokenization path. This lets us adjust the mapping from
continuous semantic features to discrete speech tokens while preserving the
encoder representation learned in Stage I.

This refinement stage combines two complementary tasks. The reconstruction
task uses a pre-warmed acoustic branch to reconstruct speech from the current
discrete tokens. An ECAPA-TDNN-style speaker encoder
~\citep{desplanques2020ecapa} extracts a hierarchy of speaker features from a
128-bin log-mel spectrogram. A Perceiver resampler compresses these features
into 32 global latents of 128 dimensions. A separate FSQ bottleneck with scalar
levels $[4,4,4,4,4,4]$ quantizes each latent, after which the complete set of
global speaker tokens is projected into a 512-dimensional conditioning vector
$g$. The semantic token features $z_{1:T}$ and the global acoustic condition
$g$ are combined by a 12-layer conditional feature decoder:
\begin{equation}
    U = D_{\mathrm{cond}}(z_{1:T}, g).
    \label{eq:codec-condition-decoder}
\end{equation}
A HiFi-GAN generator~\citep{kong2020hifigan} then reconstructs the waveform
from $U$, using upsampling factors $[10,4,4,2,2]$:
\begin{equation}
    \hat{x} = G_{\mathrm{wav}}(U, g).
    \label{eq:codec-waveform-decoder}
\end{equation}
The reconstruction branch is trained with multi-period and multi-resolution
discriminators. Its generator objective combines mel-spectrogram
reconstruction, multi-resolution STFT losses, adversarial and feature-matching
losses, and a speaker-quantization consistency loss:
\begin{equation}
    \begin{split}
    \mathcal{L}_{\mathrm{rec}}
    ={}&
    \lambda_{\mathrm{mel}}\mathcal{L}_{\mathrm{mel}}
    +\lambda_{\mathrm{STFT}}\mathcal{L}_{\mathrm{STFT}}
    +\lambda_{\mathrm{adv}}\mathcal{L}_{\mathrm{adv/FM}} \\
    &+\lambda_{\mathrm{spk}}\mathcal{L}_{\mathrm{spk}}.
    \end{split}
    \label{eq:codec-reconstruction}
\end{equation}
Here, $\mathcal{L}_{\mathrm{spk}}$ aligns the continuous and quantized speaker
representations. Reconstruction is initially conditioned on the continuous
speaker feature and subsequently switched to its quantized counterpart. This
curriculum stabilizes waveform learning before exposing the decoder to the
discrete acoustic bottleneck.

To balance the generative reconstruction signal, we also attach an adaptor and
a small LLM backbone on top of the discrete speech tokens and optimize a
multi-task speech understanding objective. This auxiliary understanding task
encourages the refined token distribution to remain useful for content-centric
speech reasoning rather than being shaped only by waveform reconstruction.

\paragraph{Unified generation interface.}
As illustrated in Figure~\ref{fig:ttsonic-overview}, the same Text-to-Semantic
Speech LM and acoustic flow-matching decoder support multiple generation modes
by composing different semantic, prosodic, and timbre conditions.
Table~\ref{tab:generation-mode-inputs} summarizes the inputs supplied to the two
modules in each mode. The language model predicts semantic--prosodic speech
tokens, while the flow-matching decoder combines these tokens with the
available acoustic or timbre conditions. Speech enhancement and voice
conversion instead bypass the language model and directly use speech tokens
extracted by the codec. For speech enhancement, tokens from the degraded input
are rendered as clean speech; for voice conversion, tokens from the source
utterance are rendered under the target-speaker condition.

\subsection{Text-to-Semantic Language Model}
\label{sec:method-text2semantic}

\paragraph{Pre-training.}
We follow the scalable token-based generation recipe established by the
CosyVoice series~\citep{du2024cosyvoice2,du2025cosyvoice3} and train an
autoregressive language model to predict the semantic--prosodic speech tokens
introduced in \cref{sec:method-codec}. We initialize the autoregressive
language-model weights from Qwen3.5~\citep{qwen3.5} and then adapt the model to
speech-token generation. Given a text sequence $y$ and its target speech-token
sequence $s=(s_1,\ldots,s_T)$, the pre-training objective is the standard
next-token negative log-likelihood
\begin{equation}
    \mathcal{L}_{\mathrm{LM}}
    =
    -\sum_{t=1}^{T}
    \log p_{\theta}(s_t \mid y,s_{<t}).
    \label{eq:text2semantic-lm-objective}
\end{equation}
Rather than introducing an isolated output space for speech, we extend the
text vocabulary with the complete set of speech tokens. Text and speech
therefore share a unified vocabulary, token embedding, and language-modeling
head. This formulation preserves the linguistic knowledge of the text
backbone while allowing lexical context, punctuation, and speech units to
interact in the same autoregressive space, improving the model's overall text
understanding and speech-token generation capability.

\subsection{Acoustic Flow Matching}
\label{sec:method-flow-matching}

\paragraph{Token-to-mel generation.}
We follow the causal conditional flow-matching design of
CosyVoice~3~\citep{du2025cosyvoice3} to convert the discrete
semantic--prosodic sequence into an acoustic representation. The model takes
the speech tokens and, when the generation mode provides one, a
192-dimensional speaker embedding extracted by CAM++~\citep{wang2023campp}
or generated by Text2Timbre. Since the tokenizer operates at 25~Hz whereas
the 80-bin mel-spectrogram is represented at 50~Hz, every speech-token feature
is repeated over two mel frames, establishing a fixed token-to-mel ratio of
$1{:}2$. A causal Diffusion Transformer then estimates the conditional vector
field~\citep{lipman2023flow} and generates the target mel-spectrogram. Finally,
a causal HiFi-GAN/HiFT vocoder~\citep{kong2020hifigan} converts the generated
mel features into a 24-kHz waveform.

\paragraph{Prompt-based cloning condition.}
In addition to the speech tokens and an optional global speaker embedding, the decoder can
optionally receive a prompt mel-spectrogram as an observed acoustic prefix. The
prompt mel is aligned with its corresponding prompt speech tokens under the
same $1{:}2$ ratio and prepended to the target generation sequence. During flow
generation, this prefix is kept as a known condition while the model predicts
the subsequent target mel frames; the prompt portion is removed before
waveform decoding. The two acoustic conditions are complementary: the CAM++
embedding specifies global speaker identity, whereas prompt mel supplies
fine-grained local acoustic evidence for higher-fidelity cloning. Omitting the
prompt mel retains a unified path for standard TTS and text-guided voice
design.

\paragraph{Training strategy.}
The acoustic model and the language model place different demands on their
training corpora. Language-model training benefits primarily from broad
coverage of languages, text forms, domains, and difficult pronunciation
patterns. In contrast, flow-matching training is more sensitive to the
fidelity of its acoustic targets. We therefore construct its training mixture
with audio quality as the primary criterion, emphasizing clean single-speaker
recordings, reliable segmentation, sufficient speech occupancy, and high
perceptual quality. We initialize the flow module from CosyVoice~3
flow-matching weights and continue training it with our speech tokens, CAM++
speaker space, and acoustic data distribution. After this supervised
flow-matching training stage, we further post-train the acoustic flow model
with FlowTTS-GRPO~\citep{wang2026flowtts} to further improve
acoustic quality, speaker similarity, and content preservation. Empirically,
the CosyVoice~3 initialization stabilizes optimization and provides a modest improvement
in speaker and timbre similarity over training the acoustic model from
scratch. During acoustic flow-matching training, speech tokens and speaker
embeddings are always extracted from the same utterance. No cross-speaker
token--speaker conditioning pairs are introduced during training.

\subsection{Instruction Following}
\label{sec:method-instruction-following}

\subsubsection{Controllable Cloning}
\label{sec:method-instruction-following-llm}

\paragraph{Problem formulation.}
Controllable cloning aims to preserve the identity of a reference speaker while
rendering the target text according to a natural-language style instruction.
Let $r$ denote the reference speech, $y$ the text to be synthesized, and $i$
the instruction. In our factorized architecture, the speaker encoder extracts
a timbre condition $g_r$ from $r$, whereas the autoregressive LLM maps $(i,y)$
to a semantic--prosodic token sequence $s$. The two conditions are combined
only by the flow-matching decoder:
\begin{equation}
    g_r = E_{\mathrm{spk}}(r),
    \qquad
    s \sim p_{\theta}(s \mid i,y),
    \qquad
    \hat{x} = F_{\phi}(s,g_r).
    \label{eq:controllable-cloning-formulation}
\end{equation}
This separation assigns complementary roles to the two inputs: the reference
speech specifies \emph{who} is speaking, while the instruction controls
\emph{how} the target content is delivered.

\paragraph{Multi-view instruction annotation.}
High-quality instruction data is essential because expressive speech admits
multiple valid textual interpretations at different levels of abstraction. We
therefore use a state-of-the-art proprietary omni model to annotate each
training utterance multiple times under two complementary protocols. The first
produces attribute-wise descriptions of perceptually grounded factors, such as
pitch, speaking rate, loudness, emotion, intonation, and rhythm. The second
produces free-form descriptions that summarize the overall speaking style,
communicative context, or role embodied by the utterance. For each view, we
retain multiple annotation realizations rather than collapsing them into a
single canonical prompt. During training, we randomly sample an annotation
view and one of its realizations for the same speech target. This stochastic
pairing exposes the model to both precise local controls and holistic,
open-ended requests, while reducing sensitivity to a particular wording or
prompt granularity.

\paragraph{CoT-guided instruction following.}
Directly mapping an unconstrained description to a long sequence of speech
tokens requires the LLM to infer the intended acoustic attributes and realize
them autoregressively in a single step. This is particularly unstable for
free-form instructions, which may express several correlated attributes
implicitly. We introduce a chain-of-thought (CoT) generation strategy that
inserts a compact style plan $c=(c_1,\ldots,c_M)$ between the instruction and
the speech tokens $s=(s_1,\ldots,s_T)$. The joint distribution is factorized
as
\begin{equation}
    \begin{split}
    p_{\theta}(c,s \mid i,y)
    ={}&
    \prod_{m=1}^{M}
    p_{\theta}(c_m \mid i,y,c_{<m}) \\
    &\times
    \prod_{t=1}^{T}
    p_{\theta}(s_t \mid i,y,c,s_{<t}).
    \end{split}
    \label{eq:controllable-cloning-cot-factorization}
\end{equation}
Accordingly, the model is optimized with next-token supervision over both the
CoT plan and the target speech sequence:
\begin{equation}
    \mathcal{L}_{\mathrm{ctrl}}
    =
    -\sum_{m=1}^{M}
    \log p_{\theta}(c_m \mid i,y,c_{<m})
    -\sum_{t=1}^{T}
    \log p_{\theta}(s_t \mid i,y,c,s_{<t}).
    \label{eq:controllable-cloning-cot-objective}
\end{equation}
We serialize the plan as text labels within a delimited
\texttt{<think>}$\ldots$\texttt{</think>} span immediately before the speech
tokens. Samples without CoT annotations remain usable through an empty
thinking span. At inference time, no oracle labels are required: the LLM first
predicts the CoT plan from the user instruction and target text, and then
continues autoregressively with speech-token generation. The resulting
train--inference consistency is intended to help the model resolve an
open-ended instruction into an explicit set of acoustic and prosodic decisions
before realizing the waveform.

\subsubsection{Voice Design}
\label{sec:method-voice-design}

\paragraph{Modular voice design.}
Benefiting from the factorized codec, instruction control over prosody and
voice design can be handled by separate modules. Controllable cloning
operates primarily on the semantic--prosodic speech tokens, whereas voice
design supplies the timbre condition required by the shared flow-matching
decoder. We therefore implement voice design as a standalone
instruction-to-speaker-embedding module. Given a textual voice description,
the module replaces the reference-audio speaker encoder with a generated
speaker embedding, without modifying either the text-to-semantic language
model or the downstream acoustic decoder. This plug-and-play interface allows
the same speech-token sequence to be rendered with a newly designed voice in
the absence of any reference recording.

\paragraph{Text-to-timbre generation.}
Our voice-design module builds on the Text-to-Timbre conditional
flow-matching framework of DeepDubbing~\citep{dai2025deepdubbing}. A frozen
Qwen3-Embedding encoder~\citep{zhang2025qwen3embedding} maps each natural
language description to a pooled and normalized sentence representation.
Conditioned on this representation, a Diffusion Transformer parameterizes a
conditional vector field that transports Gaussian noise to a 192-dimensional
speaker embedding. The model is trained with the conditional flow-matching
objective~\citep{lipman2023flow}, using speaker embeddings extracted from the
corresponding speech by CAM++~\citep{wang2023campp} as targets. Consequently,
the generated embeddings lie in the same speaker-conditioning space consumed
by our acoustic flow model. We randomly drop the text condition during
training and apply classifier-free guidance at inference, enabling controlled
stochastic sampling of multiple plausible timbres from the same description.

\paragraph{CoT-hybrid conditioning.}
Free-form descriptions can implicitly entangle several voice attributes or
leave important acoustic properties underspecified. To improve robustness, we
adopt a CoT-hybrid strategy with matched conditioning formats at training and
inference. During training, we randomly select an attribute-based description,
a holistic style description, or a role-play description for each utterance,
and append its structured CoT attribute profile to the selected text. At
inference, the instruction-following LLM first derives the CoT profile from the
user's original natural-language request; the serialized profile is then
concatenated with the original request and encoded jointly by the voice-design
module. The two components are complementary: the original description
preserves open-ended semantics and role information that support voice
diversity, while the CoT profile provides an explicit acoustic anchor that
reduces ambiguity. Together, this conditioning scheme is designed to support free-form requests
while preserving the open-ended semantics needed for voice diversity.

\section{Experiments}
\label{sec:experiments}


\subsection{Experimental Setup}

\paragraph{Benchmarks.}
\textbf{Seed-TTS-Eval} is a standard Chinese--English zero-shot TTS benchmark
with Chinese (\emph{test-zh}), English (\emph{test-en}), and challenging
Chinese (\emph{test-hard}) subsets. \textbf{CV3-Eval} evaluates multilingual
and cross-lingual zero-shot synthesis on standard Chinese and English subsets,
hard Chinese and English subsets, and cross-lingual English-to-Chinese and
Chinese-to-English subsets. \textbf{InstructTTSEval}
\citep{huang2025instructttseval} evaluates natural-language-guided speech
generation in Chinese and English through acoustic-parameter specification
(APS), descriptive-style directive (DSD), and role-play (RP) tasks.
\textbf{Seed VC} uses the English and Chinese subsets of Seed-TTS-Eval
\citep{anastassiou2024seed} to evaluate zero-shot voice conversion. Given a
source utterance and a target-speaker reference, a system should preserve the
source linguistic content while rendering it in the target voice.
\textbf{VoiceBank-DEMAND-16k}~\citep{jacoblinvoicebankdemand16k} is used to
evaluate noisy-reference voice cloning, where the system must synthesize clean
and faithful target speech from reference audio corrupted by background noise.

\paragraph{Comparison systems.}
For Seed-TTS-Eval, we group comparison systems by capability scope. The
TTS-only baselines include F5-TTS \citep{chen2024f5}, VibeVoice
\citep{peng2025vibevoice}, dots.tts(Pre.) \citep{lian2026dots},
MiniMax-Speech \citep{zhang2025minimax}, FireRedTTS2
\citep{xie2025fireredtts2}, CosyVoice~2 \citep{du2024cosyvoice2}, CosyVoice~3
\citep{du2025cosyvoice3}, Qwen3-TTS-Base \citep{hu2026qwen3}, IndexTTS2
\citep{zhou2026indextts2}, and FireRedTTS3-Base
\citep{shen2026fireredtts3}. The unified baselines, which support broader
speech generation or editing capabilities, include MiMo-Audio
\citep{zhang2025mimo}, VoxCPM2 \citep{zhou2026voxcpm2}, Auk-Flash, and Auk
\citep{ma2026auk}. For CV3-Eval, we compare with VoxCPM2, dots.tts(Pre.),
CosyVoice~3-0.5B, CosyVoice~3-1.5B, Qwen3-TTS-12Hz-1.7B-Base,
Qwen-Audio-3.0-TTS \citep{xiang2026qwen}, IndexTTS2.5, and IndexTTS2.5-RL
\citep{li2026indextts}. For InstructTTSEval, we compare with
MiMo-Audio-7B-Instruct, VoiceSculptor-VD \citep{hu2026voicesculptor},
Qwen3-TTS-12Hz-1.7B-VD, VoxCPM2, Auk, and FireRedTTS3-Instruct. For Seed VC,
we compare with three dedicated zero-shot VC systems: Seed-VC
\citep{liu2024seedvc}, MeanVC2 \citep{ma2026meanvc2}, and X-VC
\citep{zheng2026xvc}. For VoiceBank-DEMAND-16k, we compare noisy-reference
cloning against CosyVoice~3. Baseline values are taken from the corresponding
publications, official releases, or our reproduced evaluation. A dash
indicates that a value is unavailable or undisclosed.

\paragraph{Metrics.}
We evaluate content preservation with word error rate (WER) or character error
rate (CER), using Whisper-large-v3 \citep{radford2023whisper} for English and
Paraformer \citep{gao2022paraformer} for Chinese. Following the respective
comparison protocols, Seed-TTS-Eval reports CER for \emph{test-zh} and
\emph{test-hard} and WER for \emph{test-en}; CV3-Eval and Seed VC report WER
for all subsets. Speaker similarity (SIM) is the cosine similarity between
WavLM-based \citep{chen2022speaker} embeddings of the generated and reference
speech. For InstructTTSEval, we use the official evaluation protocol and
report the accuracy of APS, DSD, and RP; higher is better for all three. We
additionally use ERes2Net speaker similarity for noisy-reference cloning and
UTMOS \citep{saeki2022utmos} as a non-intrusive estimate of perceptual quality
for Seed VC and VoiceBank-DEMAND-16k. For voice-design diversity, we report the
high-similarity speaker collision rate, defined as the percentage of
within-gender--age WavLM embedding pairs whose cosine similarity exceeds a
specified threshold. Individual-pass uses each model's own judge-passing
samples, whereas Joint-pass uses samples that pass jointly for all three
systems in the diversity comparison; lower collision rates indicate greater
diversity. The objective
tokenizer ablation additionally reports DNSMOS OVRL and Spectrum MAE in dB,
while the subjective ablation reports NMOS for perceived naturalness, SMOS for
perceived speaker similarity, and CMOS for comparative preference. Lower
WER/CER, Spectrum MAE, and collision rates are better, while higher SIM,
DNSMOS, UTMOS, InstructTTSEval, NMOS, SMOS, and CMOS scores are better.


\subsection{Zero-Shot Synthesis on Seed-TTS-Eval}
\label{sec:seed-tts-eval}

\begin{table}[t]
    \centering
    \normalsize
    \setlength{\tabcolsep}{2.0pt}
    \caption{\textbf{Zero-shot synthesis on Seed-TTS-Eval.}
    Models are grouped by capability scope. TTS-only models are specialized
    for text-to-speech synthesis, whereas unified models support broader speech
    generation or editing capabilities. Among unified models, bold and
    underlined values denote the best and second-best results in each column,
    respectively. Tied best results are bolded, and the next distinct value is
    underlined. Parameter counts marked with $^{*}$ exclude the frozen
    pre-trained Qwen2.5-Omni used by the model. ``--'' denotes unavailable or
    undisclosed information.}
    \label{tab:seed-tts-eval}
    \resizebox{0.7\linewidth}{!}{
    \begin{tabular}{lllcccccc}
            \toprule
            \multirow{2}{*}{\textbf{Model}}
            & \multirow{2}{*}{\textbf{Params}}
            & \multirow{2}{*}{\textbf{Capability}}
            & \multicolumn{2}{c}{\textbf{test-zh}}
            & \multicolumn{2}{c}{\textbf{test-en}}
            & \multicolumn{2}{c}{\textbf{test-hard}} \\
            \cmidrule(lr){4-5} \cmidrule(lr){6-7} \cmidrule(lr){8-9}
            & & & CER$\downarrow$ & SIM$\uparrow$
            & WER$\downarrow$ & SIM$\uparrow$
            & CER$\downarrow$ & SIM$\uparrow$ \\
            \midrule
            Human
                & -- & -- & 1.26 & 0.755 & 2.14 & 0.734 & -- & -- \\
            \midrule
            F5-TTS \citep{chen2024f5}
                & 0.3B & \multirow{10}{*}{TTS-only}
                & 1.52 & 0.741 & 2.00 & 0.647 & 8.67 & 0.713 \\
            VibeVoice \citep{peng2025vibevoice}
                & 1.5B & & 1.16 & 0.744 & 3.04 & 0.689 & -- & -- \\
            dots.tts(Pre.) \citep{lian2026dots}
                & 2B & & 0.96 & 0.805 & 1.34 & 0.768 & 6.46 & 0.792 \\
            MiniMax-Speech \citep{zhang2025minimax}
                & -- & & 0.83 & 0.783 & 1.65 & 0.692 & -- & -- \\
            FireRedTTS2 \citep{xie2025fireredtts2}
                & 1.5B & & 1.14 & 0.732 & 1.95 & 0.665 & 8.98 & 0.703 \\
            CosyVoice~2 \citep{du2024cosyvoice2}
                & 0.5B & & 1.45 & 0.748 & 2.57 & 0.652 & 6.83 & 0.724 \\
            CosyVoice~3 \citep{du2025cosyvoice3}
                & 1.5B & & 1.12 & 0.781 & 2.22 & 0.720 & 5.83 & 0.758 \\
            Qwen3-TTS-Base \citep{hu2026qwen3}
                & 1.7B & & 1.22 & 0.770 & 1.23 & 0.717 & 6.76 & 0.748 \\
            IndexTTS2 \citep{zhou2026indextts2}
                & 1.5B & & 1.03 & 0.765 & 2.23 & 0.706 & 7.12 & 0.755 \\
            FireRedTTS3-Base \citep{shen2026fireredtts3}
                & 1.7B & & 1.01 & 0.809 & 1.64 & 0.772 & 6.50 & 0.784 \\
            \midrule
            MiMo-Audio \citep{zhang2025mimo}
                & 7B &\multirow{4}{*}{Unified} & 1.96 & - & 5.37& -& 14.14& - \\
            VoxCPM2 \citep{zhou2026voxcpm2}
                & 2B &
                & \underline{0.97} & \underline{0.795} & 1.84 & 0.753 & 8.13 & 0.753 \\
            Auk-Flash~\citep{ma2026auk}
                & 1.5B$^{*}$ & & 1.10 & \textbf{0.814} & \underline{1.03}
                & \underline{0.772} & 6.43 & \textbf{0.784} \\
            Auk \citep{ma2026auk}
                & 1.5B$^{*}$ & & 1.02 & \textbf{0.814} & \textbf{1.02}
                & \textbf{0.788} & \underline{5.91} & \underline{0.782} \\
            \midrule
            \textbf{\methodname{}}
                & 2B &Unified& \textbf{0.78} & 0.784 & 1.66 & 0.717
                & \textbf{5.64} & 0.774 \\
            \bottomrule
    \end{tabular}
    }
\end{table}

Table~\ref{tab:seed-tts-eval} reports the core zero-shot TTS performance of
models with different capability scopes. \methodname{}
achieves 0.78\% CER and 0.784 SIM on \emph{test-zh}, 1.66\% WER and 0.717 SIM
on \emph{test-en}, and 5.64\% CER and 0.774 SIM on \emph{test-hard}. These
results demonstrate that \methodname{} maintains competitive content accuracy
and reference-speaker similarity while supporting a broader unified speech
generation framework beyond TTS-only synthesis.

\subsection{Multilingual and Cross-Lingual Synthesis on CV3-Eval}
\label{sec:cv3-eval}

\begin{table}[t]
    \centering
    \scriptsize
    \setlength{\tabcolsep}{2pt}
    \renewcommand{\arraystretch}{1.08}
    \caption{\textbf{Multilingual and cross-lingual synthesis on CV3-Eval.}
    WER is reported in percent, and SIM is reported as cosine similarity.
    Across all systems, bold and underlined values denote the best and
    second-best results in each column, respectively. ``--'' denotes
    unavailable or undisclosed information.}
    \label{tab:cv3-eval}
    \resizebox{\linewidth}{!}{
        \begin{tabular}{llcccccccccccc}
            \toprule
            \multirow{2}{*}{\textbf{Model}}
            & \multirow{2}{*}{\textbf{Params}}
            & \multicolumn{2}{c}{\textbf{zh}}
            & \multicolumn{2}{c}{\textbf{en}}
            & \multicolumn{2}{c}{\textbf{hard zh}}
            & \multicolumn{2}{c}{\textbf{hard en}}
            & \multicolumn{2}{c}{\textbf{en-to-zh}}
            & \multicolumn{2}{c}{\textbf{zh-to-en}} \\
            \cmidrule(lr){3-4} \cmidrule(lr){5-6}
            \cmidrule(lr){7-8} \cmidrule(lr){9-10}
            \cmidrule(lr){11-12} \cmidrule(lr){13-14}
            & & WER$\downarrow$ & SIM$\uparrow$
            & WER$\downarrow$ & SIM$\uparrow$
            & WER$\downarrow$ & SIM$\uparrow$
            & WER$\downarrow$ & SIM$\uparrow$
            & WER$\downarrow$ & SIM$\uparrow$
            & WER$\downarrow$ & SIM$\uparrow$ \\
            \midrule
            VoxCPM2 \citep{zhou2026voxcpm2}
                & 2B
                & 3.55 & 0.750
                & 6.21 & 0.716
                & \underline{8.10} & 0.699
                & 7.48 & 0.670
                & 7.76 & --
                & 6.22 & 0.643 \\
            dots.tts(Pre.) \citep{lian2026dots}
                & 2B
                & 3.51 & - & 5.24 & - & 8.55 & - & 8.48 & -
                & 10.88 & \textbf{0.746} & 4.97 & \underline{0.719} \\
            CosyVoice~3-0.5B \citep{du2025cosyvoice3}
                & 0.5B
                & 3.89 & \underline{0.800} & 5.24 & \underline{0.742}
                & 14.25 & 0.786 & 9.04 & 0.759
                & 8.48 & 0.674 & 4.99 & 0.678 \\
            CosyVoice~3-1.5B \citep{du2025cosyvoice3}
                & 1.5B
                & 3.91 & -- & 4.99 & --
                & 9.77 & 0.785 & 10.55 & 0.761
                & 8.01 & 0.669 & 4.32 & 0.664 \\
            Qwen3-TTS-12Hz-1.7B-Base \citep{hu2026qwen3}
                & 1.7B
                & \textbf{3.09} & 0.730 & \textbf{3.68} & 0.672
                & 11.24 & 0.691 & \textbf{6.53} & 0.661
                & \underline{4.77} & -- & \underline{2.77} & -- \\
            Qwen-Audio-3.0-TTS \citep{xiang2026qwen}
                & --
                & 3.35 & -- & 4.25 & --
                & \textbf{7.44} & \underline{0.787}
                & \underline{6.71} & \underline{0.766}
                & 5.23 & -- & \textbf{2.40} & -- \\
            IndexTTS2.5 \citep{li2026indextts}
                & 0.8B
                & 4.36 & 0.771 & 5.12 & 0.681
                & -- & -- & -- & -- & -- & -- & 3.62 & 0.638 \\
            IndexTTS2.5-RL \citep{li2026indextts}
                & 0.8B
                & 3.93 & 0.779 & \underline{3.89} & 0.678
                & -- & -- & -- & -- & -- & -- & 3.55 & 0.675 \\
            \midrule
            \textbf{\methodname{}}
                & 2B
                & \underline{3.19} & \textbf{0.811}
                & 4.03 & \textbf{0.769}
                & 8.68 & \textbf{0.802}
                & 7.79 & \textbf{0.770}
                & \textbf{3.44} & \underline{0.745}
                & 4.06 & \textbf{0.729} \\
            \bottomrule
        \end{tabular}
    }
\end{table}

Table~\ref{tab:cv3-eval} reports the multilingual and cross-lingual synthesis
performance of \methodname{}. It achieves WER/SIM results
of 3.19\%/0.811 on Chinese, 4.03\%/0.769 on English, 8.68\%/0.802 on hard
Chinese, 7.79\%/0.770 on hard English, 3.44\%/0.745 on English-to-Chinese,
and 4.06\%/0.729 on Chinese-to-English. Across all compared systems,
\methodname{} obtains the highest speaker similarity on five of the six
subsets and the second-highest similarity on English-to-Chinese. It also
achieves the lowest English-to-Chinese WER and the second-lowest Chinese WER,
demonstrating competitive multilingual content accuracy together with
consistent speaker preservation across standard, challenging, and
cross-lingual settings.

\subsection{Instruction Following on InstructTTSEval}
\label{sec:instructtts-eval}

\subsubsection{Instruction-Following Performance}
\label{sec:instructtts-performance}

\begin{table}[t]
    \centering
    \scriptsize
    \setlength{\tabcolsep}{2.6pt}
    \renewcommand{\arraystretch}{1.08}
    \caption{\textbf{Instruction following on InstructTTSEval.}
    APS, DSD, and RP are reported as percentages; higher is better. Bold and
    underlined values denote the best and second-best model results in each
    column, respectively. ``--'' denotes unavailable or undisclosed
    information.}
    \label{tab:instructtts-eval}
    \resizebox{0.7\linewidth}{!}{
        \begin{tabular}{lcccccc}
            \toprule
            \multirow{2}{*}{\textbf{Model}}
            & \multicolumn{3}{c}{\textbf{InstructTTSEval-ZH}}
            & \multicolumn{3}{c}{\textbf{InstructTTSEval-EN}} \\
            \cmidrule(lr){2-4} \cmidrule(lr){5-7}
            & APS$\uparrow$ & DSD$\uparrow$ & RP$\uparrow$
            & APS$\uparrow$ & DSD$\uparrow$ & RP$\uparrow$ \\
            \midrule
            Reference Audio
                & 91.0 & 83.7 & 64.1 & 92.5 & 86.7 & 66.6 \\
            \midrule
            MiMo-Audio-7B-Instruct \citep{zhang2025mimo}
                & 75.7 & 74.3 & 61.5 & 80.6 & 77.6 & 59.5 \\
            VoiceSculptor-VD \citep{hu2026voicesculptor}
                & 75.7 & 64.7 & 61.5 & -- & -- & -- \\
            Qwen3-TTS-12Hz-1.7B-VD \citep{hu2026qwen3}
                & 85.2 & 81.1 & 65.1
                & \underline{82.9} & \underline{82.4} & 68.4 \\
            VoxCPM2 \citep{zhou2026voxcpm2}
                & 85.2 & 71.5 & 60.8
                & \textbf{84.2} & \textbf{83.2} & \underline{71.4} \\
            Auk \citep{ma2026auk}
                & 83.3 & \textbf{83.4} & 68.2 & 75.6 & 81.6 & 65.9 \\ 
            FireRedTTS3-Instruct \citep{shen2026fireredtts3}
                & \underline{85.8} & 82.0 & \textbf{69.7} & 80.7 & 82.3 & \textbf{72.0} \\
            \midrule
            \textbf{\methodname{}}
                & \textbf{88.8} & \underline{82.3} & \underline{69.4}
                & 77.6 & 78.4 & 64.4 \\
            \bottomrule
        \end{tabular}
    }
\end{table}

Table~\ref{tab:instructtts-eval} shows the instruction-following performance of \methodname{} on InstructTTSEval. On the Chinese subset,
it achieves the best APS score of 88.8 and the second-best RP score of 69.4,
while its DSD score of 82.3 remains competitive with the strongest systems.
These results demonstrate effective control of both structured acoustic
parameters and more open-ended role-play descriptions. On the English subset,
the model obtains APS, DSD, and RP scores of 77.6, 78.4, and 64.4, respectively,
but trails the strongest English baselines. Thus, the unified model provides
competitive Chinese instruction-following capability, while broader
cross-lingual instruction generalization remains an important direction for
future improvement.

\subsubsection{Voice-Design Diversity}
\label{sec:instructtts-diversity}

We evaluate timbre diversity using Chinese APS samples under two judge-passing
protocols. The Individual-pass protocol uses each model's own Gemini-passing
samples, while the Joint-pass protocol uses only samples that pass for
Qwen3-TTS, VoxCPM2, and the corresponding \methodname{} repeat. Samples are
divided into six gender--age groups, and we measure the fraction of within-group
WavLM speaker-embedding pairs whose cosine similarity is strictly above a
threshold. This Micro collision rate is lower when fewer highly similar or
near-duplicate voices are generated.

\begin{table}[t]
    \centering
    \small
    \setlength{\tabcolsep}{5pt}
    \caption{\textbf{High-similarity speaker collision rates on
    InstructTTSEval-ZH.} Results are percentages of within-group WavLM pairs
    whose cosine similarity is strictly above each threshold; lower is better.
    Individual-pass uses each model's own judge-passing samples, and Joint-pass
    uses samples jointly passing Qwen3-TTS, VoxCPM2, and the corresponding
    \methodname{} repeat.
    \methodname{} is averaged over three independent generations.}
    \label{tab:instructtts-vd-diversity}
    \resizebox{0.7\linewidth}{!}{
        \begin{tabular}{lccccc}
            \toprule
            \textbf{Model}
            & $>0.50$ & $>0.55$ & $>0.60$ & $>0.65$ & $>0.70$ \\
            \midrule
            \multicolumn{6}{l}{\textit{Individual-pass: each model's own judge-passing samples}} \\
            Qwen3-TTS-12Hz-1.7B-VD \citep{hu2026qwen3}
                & 67.68 & 52.16 & 35.29 & 19.86 & 8.65 \\
            VoxCPM2 \citep{zhou2026voxcpm2}
                & 68.32 & 52.35 & 35.40 & 20.54 & 9.46 \\
            \textbf{\methodname{}}
                & \textbf{58.77} & \textbf{43.24} & \textbf{27.99}
                & \textbf{15.28} & \textbf{6.52} \\
            \midrule
            \multicolumn{6}{l}{\textit{Joint-pass: shared judge-passing samples}} \\
            Qwen3-TTS-12Hz-1.7B-VD \citep{hu2026qwen3}
                & 67.54 & 52.23 & 35.58 & 20.16 & 8.75 \\
            VoxCPM2 \citep{zhou2026voxcpm2}
                & 67.51 & 52.01 & 35.49 & 20.92 & 9.88 \\
            \textbf{\methodname{}}
                & \textbf{62.00} & \textbf{46.31} & \textbf{30.35}
                & \textbf{16.72} & \textbf{7.22} \\
            \bottomrule
        \end{tabular}
    }
\end{table}

As shown in Table~\ref{tab:instructtts-vd-diversity}, \methodname{} has the
lowest collision rate at all five thresholds under both protocols. Under
Individual-pass, at thresholds of 0.50, 0.60, and 0.70, its rates are 58.77\%,
27.99\%, and 6.52\%, compared with 67.68\%, 35.29\%, and 8.65\% for Qwen3-TTS
and 68.32\%, 35.40\%, and 9.46\% for VoxCPM2. Under the stricter Joint-pass
protocol, the
corresponding \methodname{} rates are 62.00\%, 30.35\%, and 7.22\%, still
below Qwen3-TTS (67.54\%, 35.58\%, and 8.75\%) and VoxCPM2 (67.51\%,
35.49\%, and 9.88\%). This consistent reduction indicates fewer highly similar
voices after controlling for gender, age, and instruction compliance. The
metric reflects diversity in WavLM speaker-embedding space and does not by
itself measure audio quality, naturalness, or human preference.

Taken together, the instruction-following and collision-rate results show that
the \methodname{} combines competitive instruction adherence
with greater within-group timbre diversity.

\subsection{Voice Conversion and Noisy Cloning}
\label{sec:seed-vc}

\begin{table}[t]
    \centering
    \normalsize
    \setlength{\tabcolsep}{3.5pt}
    \caption{\textbf{Zero-shot voice conversion on Seed VC.}
    WER is reported in percent. Lower WER is better; higher SIM and UTMOS are
    better. Bold and underlined values denote the best and second-best model
    results in each column, respectively. ``--'' denotes unavailable or
    undisclosed information.}
    \label{tab:seed-vc}
    \begin{tabular}{lcccccc}
            \toprule
            \multirow{2}{*}{\textbf{Model}}
            & \multicolumn{3}{c}{\textbf{Seed-TTS test-en}}
            & \multicolumn{3}{c}{\textbf{Seed-TTS test-zh}} \\
            \cmidrule(lr){2-4} \cmidrule(lr){5-7}
            & WER$\downarrow$ & SIM$\uparrow$ & UTMOS$\uparrow$
            & WER$\downarrow$ & SIM$\uparrow$ & UTMOS$\uparrow$ \\
            \midrule
            Ground Truth
                & 1.96 & 0.09 & --   & 1.33 & 0.30 & -- \\
            \midrule
            Seed-VC small \citep{liu2024seedvc}
                & \textbf{2.71} & 0.57 & 3.37
                & 2.31 & \underline{0.74} & 2.72 \\
            MeanVC2 \citep{ma2026meanvc2}
                & 8.42 & 0.60 & \underline{3.54}
                & 5.05 & 0.73 & \underline{2.99} \\
            X-VC \citep{zheng2026xvc}
                & \underline{3.10} & \underline{0.62} & 3.26
                & \underline{1.99} & 0.73 & 2.63 \\
            \textbf{\methodname{}}
                & 3.20 & \textbf{0.67} & \textbf{4.05}
                & \textbf{1.85} & \textbf{0.78} & \textbf{3.42} \\
            \bottomrule
    \end{tabular}
\end{table}

Voice conversion evaluates whether a system can preserve the linguistic content
of the source utterance while matching the target speaker and maintaining
natural audio quality. Our VC path bypasses the language model: the speech codec
first converts the source utterance into discrete tokens, and the acoustic
decoder then synthesizes speech from these tokens conditioned on the target
speaker embedding. Notably, \methodname{} is trained without any
voice-conversion targets or paired VC supervision. During acoustic
flow-matching training, the model observes only matched token--speaker
conditions, with the speech tokens and speaker embedding extracted from the
same utterance. Voice conversion is therefore evaluated as an unseen
cross-speaker recombination setting, where the source speech tokens and target
speaker embedding originate from different speakers.

As shown in Table~\ref{tab:seed-vc}, this direct path performs particularly
well on the Chinese subset, where \methodname{} achieves the best results on
all three dimensions, with 1.85\% WER, 0.78 SIM, and 3.42 UTMOS. On the
English subset, it obtains the highest SIM of 0.67 and UTMOS of 4.05, while
its WER of 3.20\% trails Seed-VC and X-VC. Averaged equally across the two
languages, \methodname{} reaches 0.73 SIM and 3.74 UTMOS, the highest averages
among the compared systems, with an average WER of 2.53\%. These results
indicate that the direct codec-based path prioritizes speaker similarity and
perceptual quality while retaining competitive content accuracy. More
importantly, this successful zero-shot cross-speaker recombination provides
functional evidence that the learned speech representation supports the
separation of linguistic content and speaker-dependent acoustic
characteristics.

\begin{table}[t]
    \centering
    \normalsize
    \setlength{\tabcolsep}{5pt}
    \caption{\textbf{Noisy-reference cloning on VoiceBank-DEMAND-16k.}
    WER is reported in percent. Lower WER is better; higher SIM and UTMOS are
    better.}
    \label{tab:voicebank-demand-noisy-cloning}
    \begin{tabular}{lcccc}
        \toprule
        \textbf{Model}
        & \textbf{WavLM SIM$\uparrow$}
        & \textbf{ERes2Net SIM$\uparrow$}
        & \textbf{WER$\downarrow$}
        & \textbf{UTMOS$\uparrow$} \\
        \midrule
        CosyVoice~3 & 0.299 & 0.564 & 1.90 & 4.23 \\
        \textbf{\methodname{}} & \textbf{0.369} & \textbf{0.601}
            & \textbf{1.45} & \textbf{4.37} \\
        \bottomrule
    \end{tabular}
\end{table}

We further evaluate noisy-reference voice cloning on
VoiceBank-DEMAND-16k~\citep{jacoblinvoicebankdemand16k}, where the reference
audio contains background noise. As shown in
Table~\ref{tab:voicebank-demand-noisy-cloning}, \methodname{} improves both speaker-similarity metrics over CosyVoice~3, increasing
WavLM SIM from 0.299 to 0.369 and ERes2Net SIM from 0.564 to 0.601. It also
reduces WER from 1.90\% to
1.45\% and improves UTMOS from 4.23 to 4.37. These results indicate that the
\methodname{} model maintains stronger content accuracy, speaker similarity, and
perceptual quality when cloning from noisy reference speech.

\subsection{Ablation of Speech Tokenizer}
\label{sec:speech-tokenizer-ablation}

\paragraph{FSQ insertion layer.}
\begin{table}[!htbp]
    \centering
    \normalsize
    \setlength{\tabcolsep}{10pt}
    \caption{\textbf{Ablation of FSQ insertion layer in the speech tokenizer.}
    WER is reported in percent; lower is better.}
    \label{tab:speech-tokenizer-ablation}
    \begin{tabular}{lc}
        \toprule
        \textbf{FSQ Insert Layer} & \textbf{WER$\downarrow$} \\
        \midrule
        6  & 7.16 \\
        12 & 6.83 \\
        18 & \textbf{6.17} \\
        24 & 6.46 \\
        30 & 7.06 \\
        \bottomrule
    \end{tabular}
\end{table}

We first ablate the insertion depth of the FSQ bottleneck in the
E-Branchformer encoder to determine where discretization should be applied for
the speech tokenizer. As shown in
Table~\ref{tab:speech-tokenizer-ablation}, moving the FSQ module from layer 6
to layer 18 steadily reduces WER from 7.16\% to 6.17\%, indicating that later
encoder layers provide more mature semantic representations for discrete token
generation. Placing FSQ at even deeper layers then increases WER to 6.46\% at
layer 24 and 7.06\% at layer 30, suggesting that excessively late
discretization is also suboptimal. We therefore select layer 18 as the default
FSQ insertion point because it provides the best balance for extracting
semantic and prosodic information.

\paragraph{Training strategy ablation.}
To fairly evaluate the downstream generative utility of the two tokenizer
variants, we train separate acoustic decoders for the Stage-I and Stage-II
tokenizers, respectively. Both decoders share the same architecture, training
dataset, optimization settings, and training budget, differing only in the
discrete speech tokens provided by their corresponding tokenizers. Since the
two tokenizers induce different token distributions, independently adapting the
decoders avoids the distribution mismatch that would arise from directly
replacing the tokenizer of a fixed decoder. This controlled comparison
evaluates the effectiveness of each tokenizer as a discrete representation for
downstream speech generation.
\FloatBarrier

\begin{table}[!htbp]
    \centering
    \small
    \setlength{\tabcolsep}{5pt}
    \caption{\textbf{Objective comparison of the two tokenizer training
    stages on reconstruction and voice conversion.}
    WER is reported in percent, DNSMOS denotes the OVRL score, and Spectrum
    MAE is reported in dB. Lower WER and Spectrum MAE are better, while higher
    WavLM SIM and DNSMOS are better.}
    \label{tab:two-stage-tokenizer-ablation}
    \resizebox{0.82\linewidth}{!}{
        \begin{tabular}{lllcccc}
            \toprule
            \textbf{Setting} & \textbf{Language} & \textbf{Tokenizer}
            & \textbf{WER$\downarrow$}
            & \textbf{SIM$\uparrow$}
            & \textbf{DNSMOS$\uparrow$}
            & \textbf{Spectrum MAE$\downarrow$} \\
            \midrule
            \multirow{4}{*}{Reconstruction}
                & \multirow{2}{*}{English}
                & Stage I  & \textbf{2.82} & 0.687 & \textbf{3.215} & 5.812 \\
                & & Stage II & 2.94 & \textbf{0.705} & 3.209 & \textbf{5.580} \\
                \cmidrule(lr){2-7}
                & \multirow{2}{*}{Chinese}
                & Stage I  & 0.74 & 0.782 & \textbf{3.298} & 5.787 \\
                & & Stage II & \textbf{0.71} & \textbf{0.799} & 3.286
                & \textbf{5.543} \\
            \midrule
            \multirow{4}{*}{Voice Conversion}
                & \multirow{2}{*}{English}
                & Stage I  & \textbf{2.67} & \textbf{0.610}
                & \textbf{3.199} & - \\
                & & Stage II & 3.13 & 0.597 & 3.180 & - \\
                \cmidrule(lr){2-7}
                & \multirow{2}{*}{Chinese}
                & Stage I  & \textbf{0.71} & \textbf{0.759}
                & \textbf{3.316} & - \\
                & & Stage II & 0.81 & 0.751 & 3.300 & - \\
            \bottomrule
        \end{tabular}
    }
\end{table}

\begin{table}[!htbp]
    \centering
    \small
    \setlength{\tabcolsep}{9pt}
    \caption{\textbf{Subjective comparison of the two tokenizer training
    stages.} NMOS measures perceived naturalness, SMOS measures perceived
    speaker similarity, and CMOS measures comparative preference. Higher is
    better for all metrics.}
    \label{tab:two-stage-tokenizer-subjective}
    \begin{tabular}{llccc}
        \toprule
        \textbf{Setting} & \textbf{Tokenizer}
        & \textbf{NMOS$\uparrow$}
        & \textbf{SMOS$\uparrow$}
        & \textbf{CMOS$\uparrow$} \\
        \midrule
        \multirow{2}{*}{Reconstruction}
            & Stage I  & 3.94 & \textbf{4.11} & -0.12 \\
            & Stage II & \textbf{4.08} & \textbf{4.11} & \textbf{+0.12} \\
        \midrule
        \multirow{2}{*}{Voice Conversion}
            & Stage I  & 3.64 & \textbf{4.03} & -0.29 \\
            & Stage II & \textbf{3.91} & \textbf{4.03} & \textbf{+0.29} \\
        \bottomrule
    \end{tabular}
\end{table}

\FloatBarrier
Tables~\ref{tab:two-stage-tokenizer-ablation}
and~\ref{tab:two-stage-tokenizer-subjective} compare the two tokenizers under
matched decoder training conditions. On objective metrics, the Stage-I
tokenizer performs better in voice conversion, including higher speaker
similarity and pronunciation accuracy, which is consistent with its
ASR-supervised objective for learning semantic--acoustic separation. The
Stage-II tokenizer introduces modest trade-offs in objective content accuracy,
especially in voice conversion.

\begin{figure}[!htbp]
    \centering
    \includegraphics[width=0.9\linewidth]{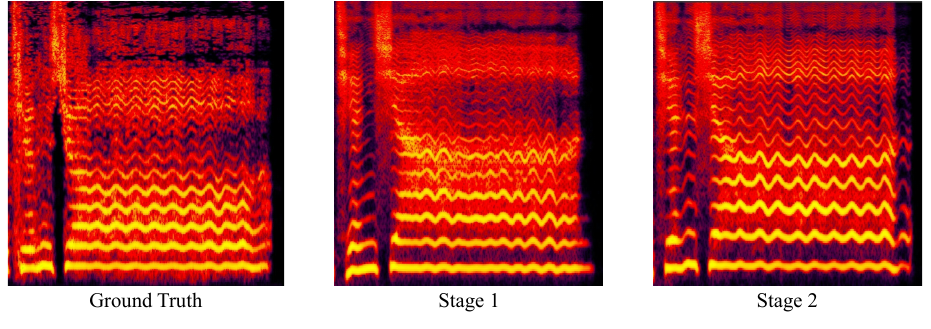}
    \caption{\textbf{Spectrogram reconstruction comparison for tokenizer
    refinement.} We compare the ground-truth spectrogram with reconstructions
    produced by the speech tokenizer after Stage-I training and after Stage-II
    refinement.}
    \label{fig:codec-reconstruction-ablation}
\end{figure}

Under the same controlled decoder training setup, however, the Stage-II
tokenizer consistently enables higher perceptual naturalness in both
reconstruction and voice conversion. Its NMOS improves from 3.94 to 4.08 in
reconstruction and from 3.64 to 3.91 in voice conversion, and CMOS also favors
Stage II in both settings, while SMOS remains comparable at 4.11 and 4.03.
These results suggest that reconstruction-guided refinement produces discrete
representations that are more favorable for downstream acoustic generation,
despite modest trade-offs in objective content accuracy. Figure~\ref{fig:codec-reconstruction-ablation}
provides a complementary visualization, where the Stage-II reconstructed
spectrogram more closely matches the ground truth.

\section{Conclusion}
\label{sec:conclusion}

In this report, we presented \methodname{}, a unified speech generation framework
that composes a semantic--acoustic disentangled speech tokenizer, a
Text-to-Semantic autoregressive language model, and an acoustic flow-matching
model. The tokenizer provides a shared discrete representation that
allows the language model to focus on semantic and prosodic planning while the
flow-matching model specializes in acoustic rendering. This factorized
framework supports zero-shot TTS, voice cloning, prosody cloning, controllable
cloning, voice conversion, and cloning from noisy reference speech. A separate
Text2Timbre module further enables flexible voice design and timbre switching.

Experiments across zero-shot, multilingual, and cross-lingual synthesis
demonstrate competitive TTS performance across the evaluated configurations.
The complete system also achieves competitive Chinese instruction following
and timbre diversity, while the voice-conversion and
noisy-reference-cloning results demonstrate effective content preservation,
speaker control, and synthesis robustness. Taken together, these results
establish \methodname{} as a practical foundation for composable, controllable,
and reliable speech generation.



\clearpage
\bibliographystyle{ieee_fullname}
\bibliography{main}


\end{document}